# WHEN WILL WE KNOW A MUON COLLIDER IS FEASIBLE? STATUS AND DIRECTIONS OF MUON ACCELERATOR R&D [*]

VLADIMIR SHILTSEV[†]

*Fermilab Accelerator Physics Center, PO Box 500, MS221*
*Batavia, IL, 60510, USA*

Over the last decade there has been significant progress in developing the concepts and technologies needed to produce, capture, accelerate and collide high intensity beams of muons. At present, a high-luminosity multi-TeV muon collider presents a viable option for the next generation lepton-lepton collider, which is believed to be needed to fully explore high energy physics in the era following the LHC discoveries. This article briefly reviews the status of the accelerator R&D, addresses the question of the feasibility of a Muon Collider , what needs to be done to prove it and presents projected timeline of the project.

## 1. Motivation and Advantages of a Muon Collider

The lifetime of the muon $\tau_0 = 2\mu s$ is just long enough to allow acceleration to high energy before the muon decays into an electron, a muon-type neutrino and an electron-type antineutrino ($\mu^- \to e^- \nu_\mu \bar{\nu}_e$). Over the last decade there has been significant progress in developing the concepts and technologies needed to produce, capture and accelerate muon beams with high intensities of the order of $O(10^{21})$ muons/year. This prepares the way for a multi-TeV Muon Collider (MC) in which $\mu^+$ and $\mu^-$ are brought to collision in a storage ring.

Muon colliders were proposed by Budker [1] in 1969 and later conceptually developed by a number of authors [2,3] and collaborations [4,5],

---

[*] This work is supported by etc, etc. This work is supported at the Fermi National Accelerator Laboratory, which is operated by the Fermi Research Association, under contract No. DE-AC02-07CH11359 with the U.S. Department of Energy.

[†] shiltsev@fnal.gov

most recently by the Neutrino Factory and Muon Collider Collaboration [6] and Fermilab Muon Collider Task Force [7]. At present, an international accelerator community works on feasibility proof of a MC needed to fully explore the physics responsible for electroweak symmetry breaking that requires a center-of-mass energy ($\sqrt{s}$) of a few TeV and a luminosity in the $10^{34}$ cm$^{-2}$s$^{-1}$ range. Figure 1 presents a layout of such a MC which has following parts: a high power **proton driver** based on "Project X" SRF-based 8 GeV H- linac [8]; pre-target accumulation and **compressor ring**(s) where very high intensity 1-3 ns long proton bunches are formed; high energy protons hit liquid mercury **target** after which muons with energy of about 200 MeV are being collected and cooled in the multi-stage **ionization cooling section** with the goal to reduce the transverse and longitudinal emittances and create a tight beam; that is followed by a **multistage acceleration (initial and main) system** – the latter employs Recirculating Linear Accelerator (RLA) to accelerate muons in a number of turns up to 2 TeV using superconducting RF technology; finally, counter-propagating muon beams are injected into a **Collider Ring** located some 100 meters underground where they stores and collide over 1000-2000 turns.

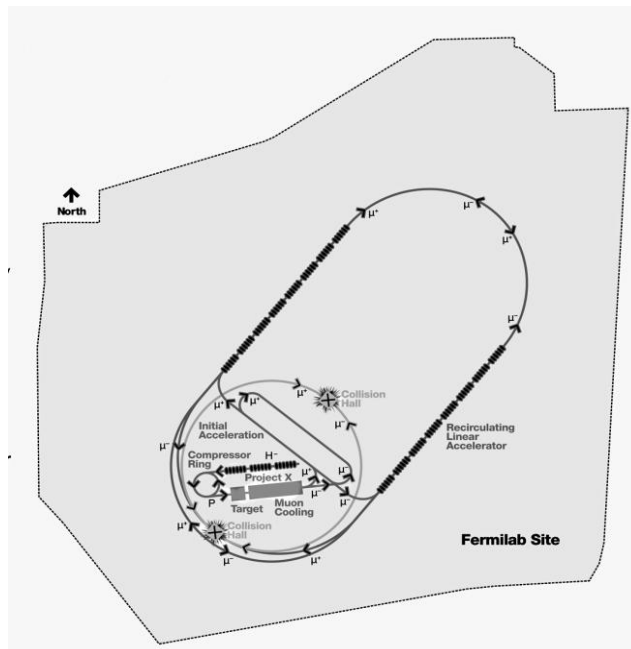

Fig. 1: Schematics for a 4 TeV Muon Collider on FNAL site.

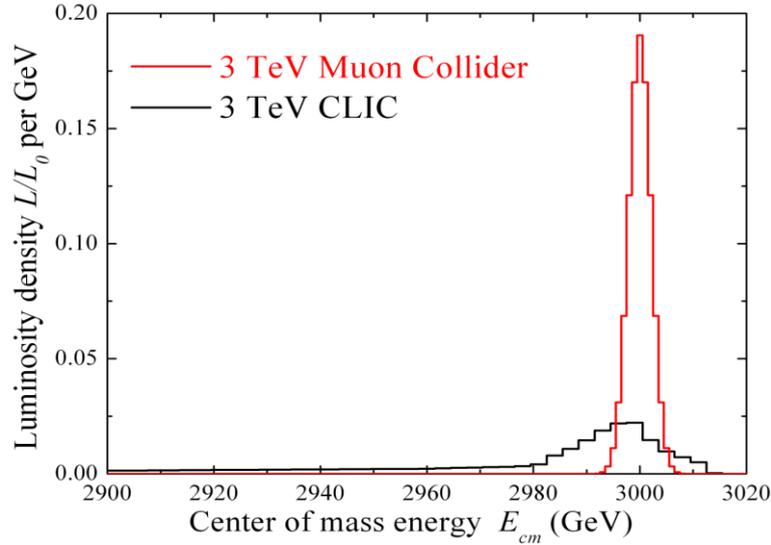

Fig. 2: Comparison of the spread of center-of-mass (com) energies for 3 TeV $\mu^+\mu^-$ collider and 3 TeV e+e- collider (CLIC).

Both e$^+$e- and $\mu^+\mu^-$ colliders have been proposed as possible candidates for a multi-TeV lepton collider to follow the Large Hadron Collider (LHC at CERN) discoveries. The physics program that could be pursued by a new lepton collider (*e+e−* or *μ+μ−*) with sufficient luminosity, would include understanding the mechanism behind mass generation and electroweak symmetry breaking; searching for, and possibly discovering, supersymmetric particles; and hunting for signs of extra spacetime dimensions and quantum gravity. By the time scale of 2014-2015, the results obtained from the LHC will be expected to more precisely establish the desired lepton collider energy.

Synchrotron radiation (proportional to the fourth power of the Lorentz factor $\gamma^4$) poses a challenge for multi-TeV e$^+$e- colliders, which cannot be circular, but must have a linear geometry and, with practical acceleration schemes, be tens of km long. Furthermore, beam-beam effects at the collision point induce the electrons and positrons to radiate, which broadens the colliding beam energy distributions. Since $(m_\mu/m_e)^4 = (207)^4 = 2\times10^9$, all of these radiation-related effects can be mitigated by using muons instead of electrons. A multi-TeV $\mu^+\mu^-$ collider can be circular and therefore have a compact geometry that will fit on existing accelerator sites (see Fig.1 for a possible footprint of MC on the 6x7 km FNAL site). The c.o.m. energy spreads for 3-TeV $e^+e^-$ Compact Linear Collider (CLIC at CERN [9]) and $\mu^+\mu^-$ collider are compared in Fig.2.

In addition to the smaller size (shorter length of the enclosures filled with high tech equipment – e.g. accelerator tunnels), smaller energy spread and potentially higher energy reach, the other advantages of the MC compared to both the International Linear Collider (ILC [10]) and CLIC are lower required wall plug power and significantly smaller number of elements which require high reliability and individual control for effective operation (see Table I). These elements are either superconducting or conventional RF structures, precise focusing quadrupoles; SC or conventional dipole magnets, etc [11]. There are a total of about 260,000 components in the 3 TeV CLIC, most of them combined in about 20,000 pre-assembled Two-Beam Accelerator (TBA) modules, each comprising of a number of accelerating and power extraction RF structures and focusing quadrupole magnets. The number of the ILC elements is about 38,000 including 17,280 SC RF cavities, 13,190 magnets (in the main linacs, damping rings, ring-to-main-linac transfer lines, beam delivery system), and some 8,000 klystrons and RF power distribution components.

The estimated number of elements of a MC is significantly less, about 10,000: that includes some 500 elements in the Proton Driver (400 cavities and 100 magnets in the main linac and transfer lines), approximately 600 magnets in the accumulator and the bunch compression rings; 1600 elements in the target and cooling sections (RF cavities, magnets, absorbers); initial and main acceleration sections based on SC RF RLA totaling about 3,000 SC RF cavities and 1,200 of magnets and klystrons; bending arcs - about or less than 3,000 magnets. For comparison, the LHC accelerator complex has total of about 11,000 elements: 9300 SC dipole, quadrupole and correct magnets, 720 magnets in the injection lines, 744 conventional magnets in the SPS and about 200 in other accelerators in the injection chain, plus about 100 RF cavities [12].

Table I: Comparison of Lepton Collider alternatives

|  | ILC | MC | CLIC |
|---|---|---|---|
| c.o.m energy, TeV | 0.5 | 1.5-4 | 3 |
| Feasibility report | 2007 | 2014-16 | 2011 |
| Cost related: |  |  |  |
| Hi-Tech length, km | 36 | 14-20 | ~60 |
| Wall plug power, MW | 230 | 120-200 | 380-430 |

Table II: The parameters of the Muon Collider options

|  | Low $E$ | High $E$ | Low $L$/High $\varepsilon$ |
|---|---|---|---|
| COM energy (TeV) | 1.5 | 4 | 2 |
| Luminosity(cm$^{-2}$s$^{-1}$) | $10^{34}$ | $4\cdot 10^{34}$ | $4\cdot 10^{30}$ |
| # of bunches | 1 | 1 | 12 |
| µ's/bunch, $10^{12}$ | 2 | 2 | 0.1 |
| Circumference, km | 3 | 8.1 | 3 |
| β* = σ$_z$ , mm | 10 | 3 | 5 |
| dp/p (rms, %) | 0.1 | 0.12 | 0.01 |
| Ring depth, m | 13 | 135 | 13 |
| PD rep rate, Hz | 12 | 6 | 60 |
| PD power, MW | ≈4 | ≈2 | 2.4 |
| Tr-emm.ε$_T$ π µmrad | 25 | 25 | 3000 |
| L-emm. ε$_L$ π mmrad | 72 | 72 | 25 |

All three lepton collider concepts may anticipate difficulty to prove feasibility of the performance - that is high luminosity of the order of the $10^{34}$ cm$^{-2}$s$^{-1}$. On the other hand, only the ILC concept can boast the full technical feasibility and readiness for construction if the high cost can be accepted. Feasibility of the two-beam acceleration scheme – the base technique of the CLIC collider – is expected to be demonstrated in 2011 [9]. The challenges of the MC are numerous (see next Section), but main condition to claim its feasibility is thought to be demonstration of the significant reduction of the 6-dimensional muon beam phase space volume (muon cooling) and resolution of the related issue of normalconducting RF cavities breakdown in strong magnetic fields. The latter is expected to be addressed by 2014-15, while convincing demonstration of the 6D cooling might take another 4 to 6 years.

To be precise, one has to distinguish between various options of the Muon Collider: a) low com energy and low luminosity collider (e.g. a Higgs factory with com energy of 100-150 GeV and luminosity in the range of $10^{31}$ cm$^{-2}$s$^{-1}$ [13]); b) high energy and low luminosity collider (eg a Z' factory with com energy of the order of 1-5 TeV but luminosity of $\geq 10^{30}$ cm$^{-2}$s$^{-1}$ sufficient to explore the new gauge bosons because of expected significant resonant of the cross section enhancement [14]); and c) high energy and high luminosity muon collider. The parameters of several MC options under study are given in the Table. The first two columns are for MCs with higher and lower c.o.m. energies and small emittances which are believed to be in principle achievable (that requires significant R&D – see next Section), while the last column is for a 2TeV MC with large beam emittances (without significant cooling) [15]. Of course, the physics reach of these options is quite different, but only the "no-cooling" option can be discussed as technically feasible at present.

Additional attraction of a MC is its possible synergy with the Neutrino Factory concept [16]. The front-end of a MC, up to and including the initial cooling channel, is similar (perhaps identical) to the corresponding Neutrino Factory (NF) front-end [17]. However, in a NF the cooling channel must reduce the transverse emittances ($\varepsilon_x, \varepsilon_y$) by only factors of a few, whereas to produce the desired luminosity, a MC cooling channel must reduce the transverse emittances (vertical and horizontal) by factors of a few hundred and reduce the longitudinal emittance $\varepsilon_L$ by a factor $O(10)$ – see Fig.3.

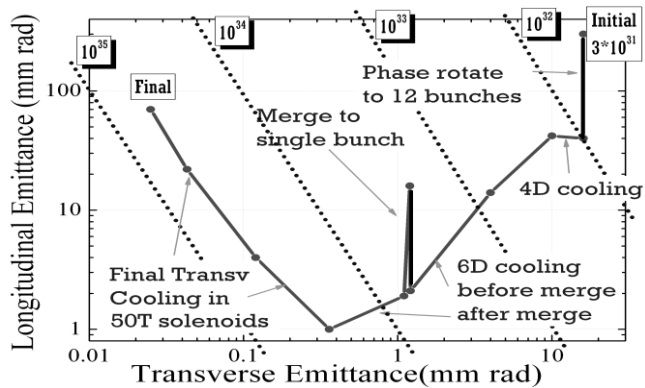

Fig. 3: Simulated 6D cooling path corresponding to one particular candidate MC cooling channel. The first part of the scheme (indicated by "4D Cooling") is identical to the present baseline NF front-end. Dashed lines indicate approximate luminosity reach of a 3TeV MC.

## 2. Recent Progress and Future Directions of Muon Collider Accelerator R&D

Muon Collider and Neutrino Factory R&D has been supported in the U.S. for the last decade and carried out by two teams of accelerator scientists associated into the US-wide Neutrino Factory and Muon Collider Collaboration (NFMCC) and Fermilab's Muon Collider task Force (MCTF). The main R&D accomplishments to date include: *a*) the construction and successful completion of an international proof-of-principle MC/NF high-power liquid mercury target experiment (MERIT); *b*) the launching of an international 4-dimensional muon ionization cooling demonstration experiment (MICE); and *c*) a series of NF design and simulation studies that have progressively improved the performance and cost-effectiveness of the simulated NF design and prepared the way for a corresponding MC end-to-end design [18].

In 2008, the Particle Physics Project Prioritization Panel (P5) has recommended *"...R&D for alternative accelerator technologies, to permit an informed choice when the lepton collider energy is established"* [19]. In response, the NFMCC and MCTF organizations are now being merged into a new national organization, US Muon Accelerator Program (MAP) which in 2010 has submitted to the DOE a proposal for a unified R&D program for the years 2010–2016 [20]. The present annual level of DOE support for all MC/NF-related R&D in the U.S. is about $10M. For comparison, this is about half of the size of the CLIC accelerator R&D program support at CERN. The requested funding for the MAP corresponds to a 50% increase in annual funding for the "nominal" profile, or up to a 90% increase for an "augmented" program that would deliver the results in less time. With this increased support we expect to demonstrate feasibility of the MC based on a credible design, an end-to-end simulation of the full accelerator complex, and an initial cost range. The main technical goals of the MAP include: (i) delivering a Muon Collider Design Feasibility Study Report (MC DFSR) - interim by the end FY14 and final (with the MC cost range) by the end of FY16; (ii) 4-Dimensional Muon Ionization Cooling Experiment (MICE at RAL, UK) completion by the end of FY13; (iii) completion of a program of RF studies to provide input for down-selection of 6D cooling channel technology by the end of FY12; (iv) participation in an International Design Study and completion of the NF Reference Design Report in early FY14; (v) construction and test of a section (unit) of a 6D muon cooling channel by the end of FY16. The proposed hardware R&D will guide, and give

confidence in, the simulation studies. The program is foreseen to comprise participants from the host U.S. laboratory (FNAL), from a number of other U.S. laboratories (ANL, BNL, Jlab, LBNL, SLAC), from universities and from hi-tech companies. Significant international collaboration with the UK, and with other countries, to understand, develop and exploit the accelerator science and technology of muon accelerators is also anticipated.

It is also anticipated that around 2014-2016, the need and feasibility of a Muon Collider will be well understood and – if the MC path found attractive for HEP community – a series of demonstration experiments with muon beams on the 6D cooling, production and collection will need to be carried out that will take (estimated) 5-7 years starting 2015-2016. In the case of success at that stage and if the HEP community wishes to go down this path, a MC construction start in the early to mid-2020s is plausible.

Prospects for a MC and/or a NF in the U.S. have recently improved due to the possibility of launching Project-X at Fermilab, since the SC proton linac could ultimately serve as the required proton driver. It is specified in the Project-X design that it has to be upgradeable from initial proton beam power of 1MW to 4MW, so it can serve as a proton source for a MC. The design work on the following accumulation/(and) bunching ring(s) has just been started recently [21].

Multi-MW target R&D has greatly advanced in recent years, and has culminated in the Mercury Intense Target experiment (MERIT [22]) which has successfully demonstrated a Hg-jet injected into a 15T solenoid and hit by an intense proton beam from the CERN PS. A high-Z target is chosen to maximize $\pi^{\pm}$ production. Solenoid radially confines essentially all $\pi^{\pm}$ coming from the target. The Hg-jet choice avoids the shock and radiation damage related target-lifetime issues that arise in a solid target. The jet was viewed by high speed cameras (Fig. 4) which enabled measurement of the jet dynamics. MERIT results suggest this technology could support beam powers in excess of 4MW.

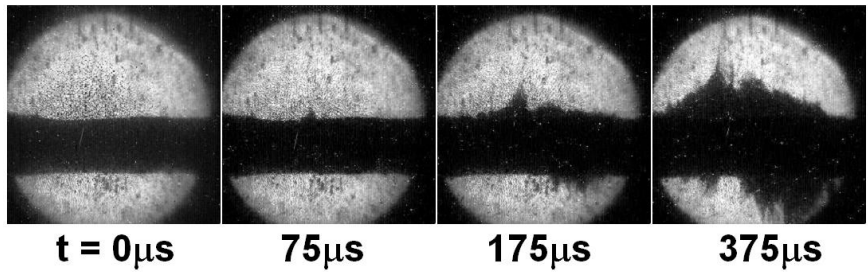

Fig. 4: Sequential images of a Hg-jet target hit by a 24GeV beam pulse containing $10^{13}$ protons (MERIT). The jet was in a 10T field (measurements have been made up to 15T). At the timescales of ~15ms the jet re-establishes itself ready for next proton pulse.

Significant efforts are presently focused on high gradient normal conducting rf cavities operating in multi-Tesla magnetic fields as required in the bunching, phase rotation, and cooling channel designs. Closed 805MHz rf cells with thin Be windows have shown significant reduction of maximum rf gradient in 3T field – 12MV/m vs 17MV/m specified. Further R&D will be part of the MAP and will explore possibilities of surface treatments, usage of high pressure hydrogen gas, "magnetically insulated" or open cavities.

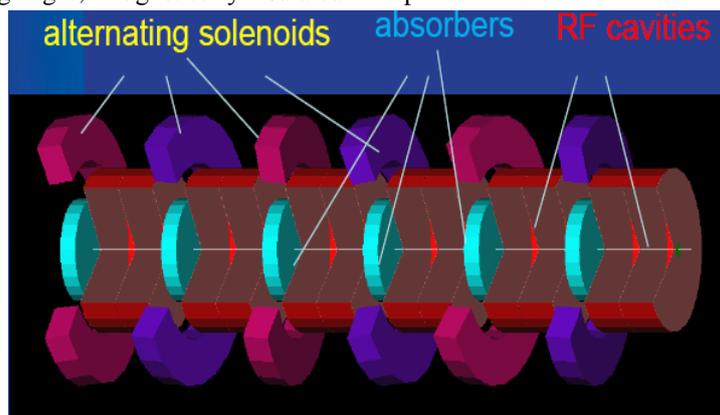

Fig. 5: Candidate scheme for 6D muon cooling ("FOFO snake") which offers fast reduction of the beam longitudinal and transverse emittances for both signs of muons.

The present baseline 4D ionization cooling channel design consists of a sequence of LiH absorbers and 201 MHz rf cavities within a lattice of solenoids that provide the required focusing. The International Muon Ionization Cooling Experiment (MICE [23]) at RAL (UK) is now at the initial stage, preparing to test an ionization cooling channel cell in a muon beam by 2013. The MICE cell is adequate for a NF.

In the last few years several self-consistent concepts based on different technologies have emerged for the MC 6D cooling channel which plays a central role in reaching high luminosity (see Fig.3). To achieve desired mixing of transverse and longitudinal degrees of freedom, the muons have to be put onto a helical trajectory, e.g. as in "FOFO-snake" [24] shown in Fig.5. The design simulations of the channels are not yet complete and the main challenges are attainment of large enough dynamic apertures, taking into account realistic magnetic fields, RF cavities and absorbers, optimization of the B-fields in RF cavities and technological complexity. The design of the final cooling stages is particularly challenging as it requires very high solenoid fields (up to ~50T have been considered). The final MC luminosity is proportional to this field. The US-MAP intends to study the viability of an HTS option for these solenoids.

A Recirculating Linac with SC RF cavities (e.g. 1.3 GHz ILC like ones) is a very attractive option for acceleration of muons from low energies in cooling sections to the energy of the experiments. It offers small lengths and low wall plug power consumption but requires small beam emittances [25].
Recently, realistic collider ring beam optics has been designed which boasts a very good dynamic aperture for about dP/P=+- 0.5% and small momentum compaction [26]. The distortions due to beam-beam interaction will need to be studied as well as practical issues of the machine-detector interface.

## 3. Facilities for Muon Accelerator R&D: Now and after 2016

At present, there are two facilities dedicated to MC accelerator R&D. The Mucool Test Area (MTA) at Fermilab has cryogenic capabilities, RF power at 201 MHz & 805 MHz, Liquid $H_2$ absorber filling capability, 5 T SC Solenoid with 30 cm bore (so, a 805 MHz Cavity fits inside) and beamline which can deliver 400MeV/c protons from Fermilab's Linac to the experimental hall of the size of about 8x20 m. The MTA facility has an established program of RF cavity studies and SC coils test for the next 4-5 years and keeping the activities there uninterrupted is critical for the success of the MICE.

The MICE facility at Rutherford Appleton Laboratory (UK) is fully occupied by the 4D ionization cooling demonstration experiment up until 2013-14 and, quite probably, beyond 2014 for a possible test of the cooling with wedge absorbers. Its ~40x12 m hall can fit all the spectrometers, two or three liquid H2 absorbers, RF cavities and bucking coils needed for the experiment. Very low intensity muon beam generated by 800 MeV protons from the ISIS accelerator hitting a target is being delivered to the MICE hall via a beamline.

The main requirements for the 6D Muon Cooling demonstration experiment facility are: (i) it has to be available after 2014, when the 6D cooling technology will be selected, and the first unit assembly and test can be started there; (ii) it should offer a low intensity muon beam for experimental studies till about 2018-20; (iii) it has to have modest incremental cost and enough space for the 6D cooling demonstration experiment; (iv) it has to be upgradable/expandable to take a medium to high intensity Project-X beam when it will become available and generate high intensity muon bunches for consecutive R&D program until early 2020's; (v) after a (major) upgrade it could be used as an operational MC or NF Front End facility.

Per Ref. [24], a 20-fold reduction of 6D muon emittance can be achieved in a 120 m of the "FOFO snake" channel - one of the possible cooling channels schemes under consideration now. About the same length is needed for a MC/NF factory front-end facility which includes a target, drift section, buncher and rotator [27]. A short bunch (ideally, shorter than 10 ns long) of $10^7$-$10^{10}$ of ~200 MeV muons every 1-10 sec should suffice for the initial beam studies.

The former KTeV experiment hall and corresponding target area fit these requirements. There is an operational 120 GeV beam line which can deliver high intensity bunches of protons from Fermilabs' Main Injector to the KTeV target area for generation of short muon bunches and consecutive studies. Plans of the post-2014 muon cooling R&D program in the KTeV hall are currently under development.

## 4. Summary

A multi-TeV muon collider presents a potentially viable option for the next generation lepton-lepton collider to fully explore the energy frontier physics in the era following the LHC discoveries. In this article we reviewed the status of the muon accelerator R&D and presented its projected directions and timeline for a decade ahead. As for any other collider project, the question of a muon collider feasibility breaks into three areas: a) technical feasibility of the critical components; b) feasibility of the design performance (luminosity); and c) feasibility of the cost. At present, only the feasibility of the performance of a and low luminosity MC (both low- and high-energy) can be claimed. Technical feasibility and cost range of a high-energy high-luminosity MC will be assessed by the end of the US-MAP program at 2016. Full proof of the MC performance feasibility is possible by around 2020 after the 6D cooling demonstration experiment.

**Acknowledgments**

I acknowledge useful discussions and valuable input from S.Geer, E.Ramberg, D.Neuffer, H.White, M.Syphers, R.Abrams M.Zisman, A.Tollestrup, A.Bross, Y.Mori, K.Yonehara, A.Skrinsky, A.Jansson, H.Kirk, R.Palmer, Yu.Alexahin, S.Holmes, R.Johnson, D.Kaplan, D.Neuffer, Y.Derbenev, E.Eichten, R.Fernow, V.Lebedev, M.Popovic, J.Norem, M.Lamm, P.Snopok, C.Ankenbrandt, N.Mokhov, D.Summers, J.P.Delahaye, G.Geschonke, M.Chung, V.Balbekov, A.Zlobin, C.Hill, and M.Demarteau.